# Online Poker and Rummy - Games of Skill or Chance?

Taranjit Kaur, Manas Pati Tripathi, Ashirbad Samantaray, Tapan K. Gandhi

*Dept. of Electrical Engineering, Indian Institute of Technology, Hauz Khas, 110016, New Delhi*

**Abstract:** The paper aims to investigate the degree of cognitive skills required for success in online versions of the popular card game rummy and poker. The study focuses on analyzing the impact of experience and learnable skills on success in the online card game. We also propose a framework to analyze online games to conclude on whether they are games of learnable skill or are they games of chance. The hypotheses proposed aim to test whether online and offline card games are comparable in terms of cognitive engagement and skill requirements. To assess these hypotheses, key elements of gameplay such as shuffling of cards, card deck randomness, and seating of players are analyzed. We also adopted statistical approaches to understand the characteristics of card games in terms of "random chance" or skill. From the analysis, we could see that the normality of the derived variables deviates significantly from the normal distribution showing a non-linear trend. It signifies that the mean of the involved skill variables is not zero as the user plays a greater number of games, thereby strengthening the assumption that the long-term success in online card games is attributed to skill and not chance. There is no difference in online and offline versions of card games (rummy and poker) from the perspective of requirement of skills. Moreover, our finding suggests that there is a preponderance of skills to succeed in online card gaming. Overall, the findings of this research contribute to a better understanding of cognitive skills in online gaming environments.

## 1. Introduction

In the ever-evolving world of digital entertainment, online games have emerged as a captivating force that transcends borders and brings people together in virtual realms. With the rise of internet connectivity and technological advancements, online gaming experiences have become vivid in nature. Online card games like poker and rummy have established themselves as timeless classics that continue to captivate players of all skill levels. The effect of this online gaming on the sociological and psychological paradigm of people has been studied well[1]–[3].

However, there is a dearth of research on whether these games are intrinsically different than offline versions. In this study, we shall analyze if online card games, particularly rummy and poker, require similar skills to succeed or if there is more randomness and success is less dependent on skill than luck.

The brain is responsible for a large variety of functions that aid us in performing tasks like sensory perception, higher-order thinking, planning, and motor coordination[4] . These tasks are possible by means of specific specializations in the brain related to attention, memory formation, retention, recall, high-order cognitive processing, sensory association, and more. Complex interplay of these abilities gives rise to skill in particular tasks[5].

These skills are actively put to the test in human society by means of games, sports, and competitions. While games like soccer, cricket, or tennis test the players' physical, motor coordination, and motor planning abilities, card games, puzzles, or board games are known to test the players' cognitive abilities, memory, attention, and social information processing skills[6]

Classical forms of these games involve direct interaction with fellow players, game equipment, and surroundings which provides a rich flux of stimulation to the sensory, social, and cognitive processing regions of the brain [7].

In addition to proficiency in gaming tasks (like batting or bowling in cricket), the players also need to be adept in reading social cues, teamwork, etc. [8]. This elicits the question of whether the aforementioned skills indeed play a role in a player's success in this format of the game. In addition to the cognitive skills involved in these games, the results of certain games also depend on chance to a certain degree. For example, games that involve rolling die or shuffling of cards may be categorized in this set with mixed influence of skill and chance. Targeted experiments to study neural correlates of various strategies in these games need to be conducted to estimate the extent of skill they require to win. There is an ongoing debate on the contribution of chance and skill in one's performance in card games [10]. Several observational studies have attempted to test the contribution of skill to the outcome of poker games. In a study by Potter van Loon et al., [11] it was observed that players with higher performance in the earlier stages of gaming showed persistent winning in future games on the time scale of months. It was also observed that players in the bottom deciles of performance in the early stages were likely to perform badly in the later stages as well. These observations led the authors to comment that the outcomes in poker may be dominated by skill. Another article discussed that games like poker, where chance and skill both have significant contributions, would exhibit a Critical Repetition Frequency (CRF) as a threshold over which the game's outcome would be dominated by skill as the components of chance cancel out [12]. A report by Levitt and colleagues argued that the dominance of either skill or chance needs to be quantified instead of using qualitative measures in such comparative studies[13]. In an attempt to mathematically derive the influence of skill or chance in poker, another study used a basic model of the game to illustrate that a hypothetical player operating at random would be outperformed by a player that could estimate the "rank" of their hand amongst all possible hands [14]. This ability to estimate one's hand's rank was attributed to a player's skill at poker. In a controlled experimental study by Meyer et al., [15] the authors controlled the randomness of dealing cards by using three classes of hands (winner's hands, neutral hands, loser's hands) dealt to expert and average players. Both categories of players were dealt all types of hands, and it was observed that the expert players did not outperform the average players in terms of final earnings, rather the outcomes were dependent on the hand they were dealt. These results indicate the earning outcomes of poker players may significantly rely on the random hand they were dealt. However, additional analyses of the data did show that expert players minimized their losses better than average players, hinting at some involvement of restraint as a skill. In an article by Turner[16], it was discussed that poker is never entirely a game of skill or chance. It was reported that the outcome of a game was dependent only when the level of skill differs between individuals. On the other hand, the outcomes were dictated by pure chance when player's skill levels are matched. Overall, the ongoing debate has presented evidence for contribution of skill as well as chance in the outcomes of card games like poker. Conclusive remarks on this may be made upon investigating the game in a highly controlled manner with adequate markers of performance and skill. Therefore, here, we begin with review the factors that contribute to skill and their relationship with card games.

One of the important skills that is involved in bluffing- based card games like poker is the ability to navigate social and emotional cues of the players in the games including self [17]. Firstly, such games demand the ability to suppress any display of emotional information that may betray the position in the game. This suppression of emotion is essential to the game and needs to be done in tandem with fast information processing of newly handed cards. In the game, this ability is termed the "poker face". Studies have been conducted to find the neural correlates of the poker face, and to study whether the differences between a seasoned poker player and an amateur can be observed at the neural level [18]. It has been reported that brain regions including the right inferior frontal gyrus, bilateral putamen, presupplementary motor area and right supramarginal gyrus show activity when subjects are instructed to suppress their emotional responses to high-arousal stimuli [18]. Displaying such abilities during a game of poker may be linked to high activity in these regions. Regular poker players have also been reported to show superior performance in visuospatial attention in the context of social information processing from faces [19], which suggests a heightened ability to judge the opponents in a game leading to better performance overall. In this recent study on nondisordered poker players, Hurel et al. [19] used the Posner cueing paradigm to investigate the allocation of attention of these players towards social stimuli. The group of poker players showed enhanced inhibition capabilities as compared to control players. They also reported that seasoned poker players show better visuospatial attention abilities as compared to novice players. This shows that seasoned poker players exhibit greater restraint in displaying their emotions even after emotional processing is complete. This skill may be linked to the usage of the "poker face" in the game.

While playing card games in an offline setting involves rapid social cue processing, this is missing the new online formats of card games. Here, the players for a game are randomly selected from a pool of active users and the game is conducted without any facial or social information input to the players. This format necessitates the utilization of other aspects of skills which involve real time decision making based only the players' own cards and the cards that have already been played in the game. In such scenarios, pattern recognition abilities along with working memory may play a vital role in memorizing the cards in the game and playing accordingly. Moreover, pattern recognition skill is also dependent on the working memory capacity, episodic memory, focused attention, and semantic processing[20]

In a study by Schiavella and colleagues[21], it was reported that subjects who played online poker showed high correlation between quality of playing strategies and executive function. Moreover, it was observed in another investigation that working memory performance was an important predictor of performance in poker, irrespective of the players' knowledge about the game [22]. These results suggest that expert players of card games exhibit enhanced ability to recognise patterns. The relationship between performance in card games and cognitive abilities has also been studied extensively. In a study on older adults, it was found that certain moves in card games were a significant biomarker of Mild Cognitive Impairment (MCI) indicating a link between cognitive ability and card game strategies[23] . This study highlights the role of cognitive skill in playing card games and can be extended to a population of card players to study the link between cognitive ability and gaming performance. While MCI is a neurological disorder characterized by loss of memory, attention and association ability, healthy controls may also exhibit a gradation in these brain functions which could reflect in their card game strategies. Such individual differences in cognitive ability have been linked to differences in working memory, attention, and semantic processing functions[23]. Poker players have been reported to show superior performance in these domains suggesting a possible impact of a player's cognitive ability in card game outcome[21], [22]. A significant correlation between the aforementioned cognitive abilities and the winning probability of a player may be useful to test for the involvement of skill in card games.

In addition to the direct impact of memory, attention, and cognitive superiority on the strategies used in card games, these aspects are also known to improve with experience in gaming [5]. Cognitive skill acquisition involves the improvement of mental faculties that result in enhanced performance of an individual with experience [5]. Acquisition of these skills is decoupled from physical abilities. Regular training of older adults with cognitively challenging card games has been shown to improve their executive function, learning, and episodic memory [24]. The study also reported that cognitive skills showed improvement in subjects that played card games regularly as compared to casual non-frequent players [24]. These findings suggest that there is a bidirectional relationship between cognitive function and experience playing card games. Extended exposure to challenging card games may lead to enhancement in cognitive abilities, while also enabling the player to perform better.

Stafford et al. [25] have explored and analysed the learnable skills and their effect on success in online games. Although their analysis is limited to a particular game (Destiny), the framework of analysis is relevant to our analysis. In this work, researchers analyze the mean average winnings of different classes of players and made inferences. They claim that the learning curve follows the power law of learning [26]. However, Heath et al. [27] devised that exponential function could fit learning with experience curve better. In our work, we shall analyze if a learning curve exists in online poker and rummy games. At the outset, let us first state the scope of our analysis as follows. We shall restrict our analysis to,

1. Given the performance of players over a significant time period whether there is any improvement in the skill variables or not.
2. Our analysis does not analyze the effect of other possible randomness in the game such as algorithms to choose the opponent, random algorithms that shuffle the cards, and so on.
3. The effect of the awards and punishment for winning and losing the game will not be analyzed.

With this scope of the study, we hypothesize that a) the success in online card games like poker and rummy have characteristics like offline games and there is a learning curve; b) in terms of its impact on how humans engage, online and offline rummy can be considered at par, in terms of cognitive behavior and skill requirements for success in the game; c) while there are technical and operational differences between the online and offline versions of rummy, this does not translate to a reduction in the requirement for skill in online rummy as compared to offline rummy.

Contributions in this paper are as follow:

1. Explored what are the technical and operational differences in the online and the offline version of card games
2. Investigated which skill variables contribute the long-term success in online game of poker and rummy
3. We extensively analyzed the variation in these skill variables over the number of games
4. Analyzed the learning experience in different game variants

The organization of the rest of the paper is as follows. Section 2 explains the preliminary knowledge required to understand the rest of the analysis. Section 3 details the methodology of the research presented. Results and discussion have been drawn in section 4 and Section 5 concluded the work.

## 2. Preliminaries Details
2.1. Hypothesis and Constraints

Considering the scope of our study and the preliminary hypotheses formulated in the previous section, overall, the hypothesis investigated in this report is that "*The cognitive behavior and the skill in online and offline card games improves with experience*".

In order to be able to assess the above hypothesis, we first need to look at the factors that could possibly give rise to differences between offline and online card games in terms of user experience, performance or technical implementation and gameplay. Here, we have explored such differentiating elements of online and offline gameplay and highlighted the ones that will be considered given the scope of our study.

2.1.1. Shuffling of cards

In online card games, the game interface uses certified random-number-generator (RNG) software to ensure that cards are shuffled and allocated randomly, without discernible patterns in card distribution. In offline card games, game participants or a game host may be responsible for shuffling cards and ensuring that cards are randomly allocated to players. While this is an important difference between the two modes, in India, it is required that permissible online real money games (POMRGs) follow a co-regulatory framework with registered self-regulatory bodies (SRBs) which verify the authenticity of the company's random number generators [28]. Therefore, as the online gaming platforms studied in this report are POMRGs, we have excluded the analysis of RNGs from the scope of this study.

2.1.2. Card deck

In online card games, the presence of certified RNG software ensures that cards picked from the card deck are random in nature, without preferences being given to any of the players. In offline card games, game participants or a game host may be responsible for shuffling the card deck to ensure that cards are picked randomly from the deck. As discussed in Section 2.1.1, the analysis of RNGs has been excluded from the scope of this study.

2.1.3. Seating of players:

In online card games, players' seats are allocated randomly, without any prefixed seats for any players, ensuring that players have no control over the selection of players at the table or their position and seating at the table. Online card games may also ensure that multiple players from the same GPS location may not be seated at the same game table to prevent collusive gaming practices. In offline card games, ensuring randomness of seat allocations, and preventing collusiveness will depend on the game participants or game hosts. As per global best practices, POMRGs shall obtain a "no-bot" certificate that ensures all participants in a card game are real people and not company robots [34]. Therefore, we excluded this factor from our scope.

2.1.4. Access to information

In online card games, players cannot access information relating to the cards held by opponent players. Information about playing cards is encrypted to prevent third parties from viewing the same, and only a player will have information about the cards assigned to him/her. This ensures that game players use memorization skills in relation to the fall of cards held and discarded to ensure a successful outcome. In offline card games, the participants are responsible for ensuring that opponent players cannot access information about the cards held by the player.

2.1.5. Identity of players

In online card games, while players may not be visible to each other, the use of comprehensive know-your-customer processes by the game interface ensures that each player is playing against a randomly allocated human player (who has attained the age of 18 years). Online card games also undergo a no-bot certification process that works to ensure that only human players are playing against each other. In the context of offline card games, players can see and identify their opponents while playing.

2.1.6. Time constraints

Online card games might specify time constraints for game-play, which might require players to employ faster decision-making skills. Offline card games may not require this element of skill to be present.

2.2. Skill in Online Games

2.2.1. Game of Online Poker

The world of traditional poker has been completely transformed by the dynamic and quickly expanding digital gaming phenomena known as online poker. It includes a variety of card games played online using specialized poker platforms, including Texas Hold'em, Omaha, and Seven-Card Stud. In online poker, players from all over the world can virtually connect, overcoming distance and fostering a worldwide player pool [29]. Complex software that manages operations like card dealing, chip handling, and calculating winnings facilitates the gameplay. Players can take part in tournaments with varied entrance fees and prize pools or cash games where actual money is on the line [30]. Online poker's accessibility and convenience let players play their preferred games whenever they choose, from the comfort of their own homes [31]. Additionally, a vast range of game types, stake levels, and tournament formats are available on online poker sites, catering to the tastes and skill levels of a diverse player base.

2.2.2. Game of Online Rummy

Online rummy is a well-liked digital rendition of the classic card game of rummy that has become quite famous in the online gaming industry [32]. To take part in virtual games, participants link to specialised rummy platforms through the internet. The basic principles of online rummy are the same as those of its offline cousin, where players try to meld and discard their cards in order to create legitimate sets or sequences.

Players can participate in a variety of rummy variants on the digital platforms, including deals rummy, points rummy, and pool rummy, among others, with ease and an intuitive user interface. The ease of playing rummy online from the comfort of their homes or while on the go using mobile applications is available to players. Modern software that guarantees fair gameplay, random card distribution, and safe transactions makes playing online rummy a pleasant experience. These platforms frequently include interactive elements like chat features and multi-player modes that let users communicate with one another and compete against others from different backgrounds. In order to improve the game experience and create a lively player community, online rummy platforms also use a variety of player engagement methods, such as awarding loyalty programmes, bonuses, and tournaments. For new players to develop their skills and understand the complexities of the game, they also offer thorough tutorials and practice tables. It is important to remember that in the world of online rummy, responsible gaming behavior and respect to relevant laws and regulations are critical. To provide a secure and engaging gaming environment, fair play, secure transactions, and the security of personal information must be prioritized by both players and platforms. For players, industry stakeholders, and policymakers to successfully navigate this digital gaming arena and promote responsible gaming practices, it is imperative that they have a solid understanding of the mechanics and ramifications of online rummy [33].

**Methodology**

The detailed methodology is given below. The data used in this study was shared with us by All India Gaming Federation (AIGF). Our findings are based on the analysis of the data available to us.

3.1. Skill Variables in Online Poker

Three variables that contribute to overall skill in online games have been studied in this manuscript.

The big blind is a mandatory pre-flop contribution to the pot, typically twice the value of the small blind, placed by the player two seats to the left of the dealer. This blind is part of the game's structure to create initial pot value.

The concept of "big blind won" (BB/100) is a standard metric in poker analysis. It refers to the average number of big blinds won over 100 hands. By normalizing the measurement of winnings to the value of the big blind, this metric offers an objective evaluation of a player's performance. It facilitates a coherent comparison across games with different stake levels and varying blind structures.

3.1.1 Average blind amount won

It refers to the average amount of money won by the player over the number of games they play.

3.1.2 Average blind amount lost

It refers to the average amount of money lost by the player over the number of games they play.

3.1.3 Tightness

It is the ability of the players to exhibit self-control and restrict themselves from investing in a large number of hands.

3.2. Skill Variables in Online Rummy

There are three variables contributing to overall objective in a rummy game that have been studied here.

3.2.1. Average points lost by the opponent

This is dependent on the skill of the opponent (Player B in Table 1) – when the opponent is highly skilled, he loses less points (indicated as 'Low' in the Table 1) while when the opponent is low on skill, he concedes more points (indicated as 'High'). Player A has no control over this as he is randomly matched with an opponent. It can be assumed that if player is playing on a large platform with good number of players this factor will average out over a set of games (player will be matched with a high skilled player in some cases while low skilled player in some others averaging out outcome over a series of games).

3.2.2. Win Rate

The win rate is dependent on relative skill between the player and his opponent. If he is higher in skill than the opponent, then the win rate will be high. If his opponent is more skilled than him his win rate will be low. If they are equally skilled then the win rate will lie in mid-range.

3.3.3. Average Points lost in losing hands

This is solely dependent on the player's own skill. If a player is high on skill, he will lose less points. Whereas if he is low on skill, he will lose more points. This is the component of the net margin which is completely dependent on the player's own skill. With the above reasoning average points lost in losing hand comes out to be key determinant of game outcome and players own skill.

It can be said that net margin in a series of rummy deals will largely depend on the player's skill in conceding less points in the games when he has less favorable cards.

**Table I Interrelation between the skill variables**

| Player A | High Skill | | Low Skill | |
|---|---|---|---|---|
| Player B | High Skill | Low Skill | High Skill | Low Skill |
| Avg. Points Lost by opponent | Low | High | Low | High |
| Win Rate | Mid | High | Low | Mid |
| Avg. Points lost in losing hands | Low | Low | High | High |
| Net Win Margin | Mid | High | Low | Mid |

## 3.4 Data and Analysis

### 3.4.1 Poker Data

We obtained raw log files describing the action for each hand played at the 6-Player(6P) table The raw data has the following parameters:

1. *user_id*: Unique identifier for the user
2. *game_id*: *Game_id* or *game_key* is a unique identifier for the game
3. *Game Type*: Specifies whether the poker session is a "Ring" game (played with direct currency value chips) or a "Tournament" (players buy-in to compete and win prizes based on rank placement).
4. *Game Variant*: The specific set of poker rules in play, such as "Texas Hold'em" (players are dealt two private cards) or "PLO" (Pot Limit Omaha, players are dealt four private cards but must use two).
5. *Big Blind*: Mandatory chips placed before cards are dealt, used to initiate action and create a pot.
6. *Big Blind Won*: A performance metric indicating the average number of big blinds won over a given number of hands, often per 100 hands.
7. *No of Players*: The current number of active participants in the game.
8. *Max No of Players Allowed*: The maximum capacity of players that can participate in a given game or table.
9. *Min No of Players Required*: The least number of players needed for the game to commence or continue.
10. Chips placed by the player in the hand
11. Chips won by the player in the hand
12. *Game Start Date*: The specific date and time when the poker game or session commenced.
13. *Game End Date*: The specific date and time when the poker game or session concluded.
14. *Game Start Date*: The specific date and time when the poker game or session commenced.
15. Game End Date: The specific date and time when the poker game or session concluded. Some additional parameters that were considered for raw data processing are:
16. *games_played*: Total number of games played
17. *game_won*: Total number of games won
18. *Month-wise win bb/100*: A performance metric indicating the average number of big blinds won per 100 hands, segmented and analyzed on a monthly basis.
19. *VPIP of Users*: "VPIP" stands for Voluntarily Put Money in Pot, a statistic representing the percentage of hands in which a player puts chips into the pot pre-flop.

The derived set of variables resulting from the above parameters are:
1. Probability of win: It is the ratio between the number of games won by the user to the number of the games played by the user at a particular instance.
2. Mean probability for 6P scenario: It is the summation of the winning probability to the total number of games played by the user.
3. Standard Deviation in 6P scenario: The standard deviation indicates how far are the current values of the winning probability (skill variable) from the mean value.
4. Rank Average in 6P scenario: In rank average approach, it basically assigns repetitive figures to values at the same position.
5. Theoretical Quintiles in 6P scenario: It returns the inverse of the standard normal cumulative distribution. The distribution has a mean of zero and a standard deviation of one. Given the probability that a variable is within a certain distance of the mean, it calculates the z value (standard normal deviate), corresponding to an area under the curve. It is calculated using the NORM.S.INV function in excel
6. Observed Quintiles in 6P scenario: It is calculated using the standardize function in excel. The STANDARDIZE function returns a normalized value from a distribution characterized by mean and standard deviation
7. Percentile of win probabilities (95%): The percentiles are calculated using the formula of (Rank-0.5)/number of games played.

3.4.2 Rummy Data

The rummy data provided by All India Gaming Federation (AIGF) comes with a cut-off of 30 or 100 games per user in a 2P, 3P, and 6P scenario. The raw data has the following attributes:

1. *user_id*: Unique identifier for the user
2. *game_id:* Game_id / game_key is a unique identifier for the game
3. *game_type*: Points / Pool / Deal
4. *game_variant*: Value per point in case of points game, else buy in value for that game variant
5. *max_players*: Number of players for the particular game variant
6. *actual Players*: Number of players who joined the deal/game
7. *game_start:* Time at which the game started, based on the min update time in ledger for that game
8. *game_end*: Time at which the game ended
9. *deal_end*: Time at which the deal ended
10. *deal start*: Time at which the deal started
11. *buy in*: Total buy in amount for the user for that game

12. *win_amt*: Settlement transaction amount for the winner
13. *deal_id*: Unique identifier for a deal in a game
14. *deal_number*: Deal Sequence in a game
15. *Is_winnner:* Flag to indicate if a user is a winner of a deal
16. *Winner Points*: Points won by the user in a deal
17. *Loss Points*: Points lost by a user in a deal

The derived set of variables are same as that of the game of Poker in each 2P, 3P, and 6P scenario.

3.4.3 Type of Analysis

3.4.3.1 Persistence of Skill

To analyze the skill persistency, we analyzed a sample of players with minimum and maximum number games played from 30 to 100 and calculated the correlation coefficient for the skill variables.

3.4.3.2 Learning Curve Test

This test focuses on the progression of skill variables as players participate in more games. If these skill variables show improvement as the number of played games increases, it would indicate that gaining experience leads to better performance. In essence, this test looks at whether a player's performance improves with experience, suggesting that the game requires learnable skills.

3.4.3.3 Test for Normality

This test is used to check the distribution of win percentages of players. QQ plots are utilized to validate the analysis across different quintiles. If there is a significant deviation from a normal distribution, it suggests that long-term success in the game is more related to a player's skill than to randomness. This method helps in determining if consistent wins over time are more attributed to skill development or if they're merely a result of chance.

## 4. Results

4.1 Poker

In a game of pure chance, the win rate of a user in different time periods is mutually independent of each other. But in a game of skill, the win rate of a user in first time period will be positively correlated with the win rate of the user in the subsequent time period. As an example, we took all users who have played at least 30 games in both the time periods, let's say Dec 2022 and Jan 2023. Now we measured the correlation between the win rate of the users in the two time periods. The correlation came out to be 0.904, i.e., positive which indicates the persistence of skill in the game. This revelation underscores the importance of skill over mere chance in influencing outcomes over consecutive time periods. This consistency in win rates highlights the fact that mastering the nuances and strategies of the game can significantly improve one's performance. Such a high correlation suggests that players who invest time and effort in understanding the game's intricacies stand a better chance of maintaining a higher win rate. The same is being graphically illustrated in Figure 1.

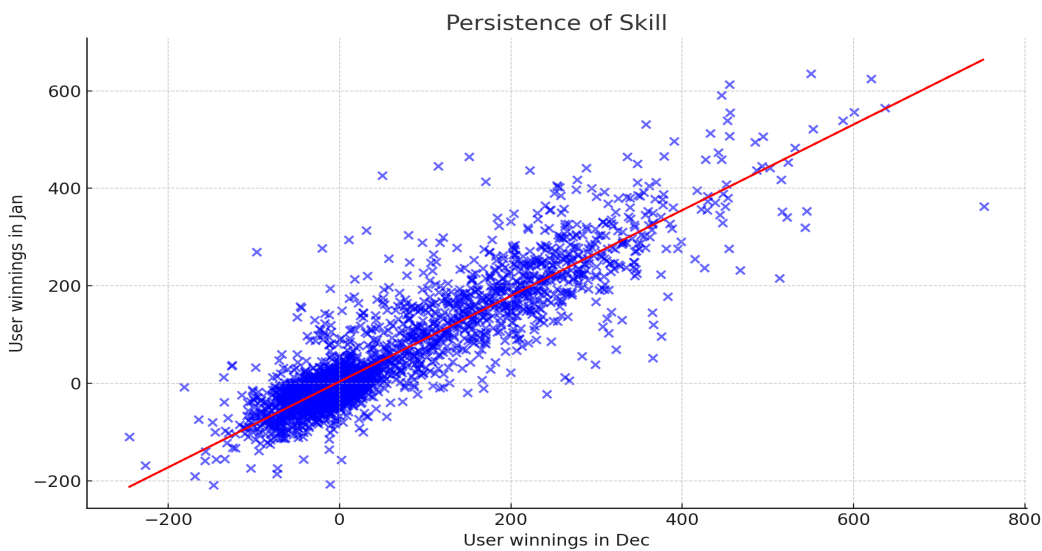

**Figure 1:** Persistence of Skill in Online Poker

We also investigated the impact of experience on learning the game of poker and the results are given in Figure 2. It demonstrates that as users engage in more games, their chances of losing decrease, and their win rate gradually improves. The graph elucidates a clear trend: as players immerse themselves in more rounds of the game, they exhibit a declining propensity to lose. This suggests a direct correlation between the number of games played and the enhancement of a player's skills and strategies. It's not just about the cards one is dealt, but how one plays them, and this proficiency seems to be honed over time and with consistent play.

The X-axis represents the number of games played by the users, with each unit movement indicating 10 sessions played. On the other hand, the Y-axis represents the average big blinds won by the users.

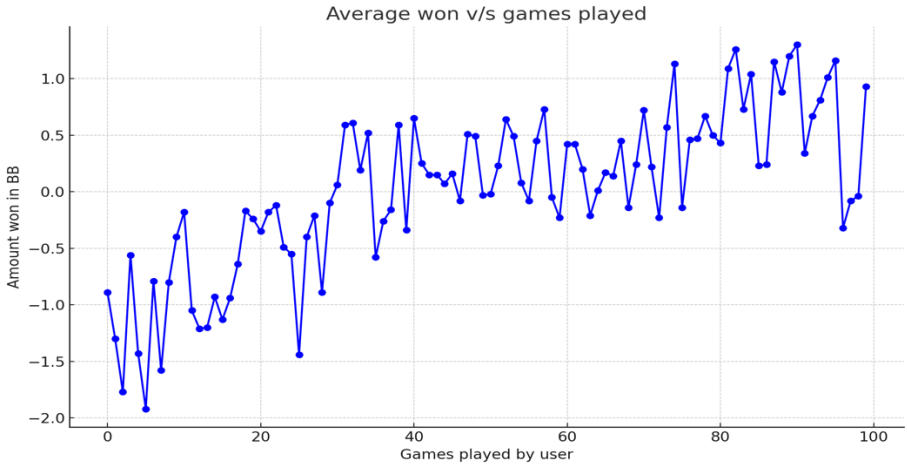

**Figure 2:** Average big blind amount won by the user vs number of games played by user in 6P scenario

We also analyzed the average points lost by the losing hands versus the number of games played by the user and results are given in Figure 3. It indicates a decrease in the amount lost with the number of games played. It is evident that as users embark on their poker journey, they are more prone to incurring losses. However, as experience accumulates, their losses gradually diminish. The X-axis represents the number of games played by the users, with each unit denoting 10 sessions. Conversely, the Y-axis signifies the average big blinds lost by users in losing hands, shedding light on the improvement in their strategic play over time.

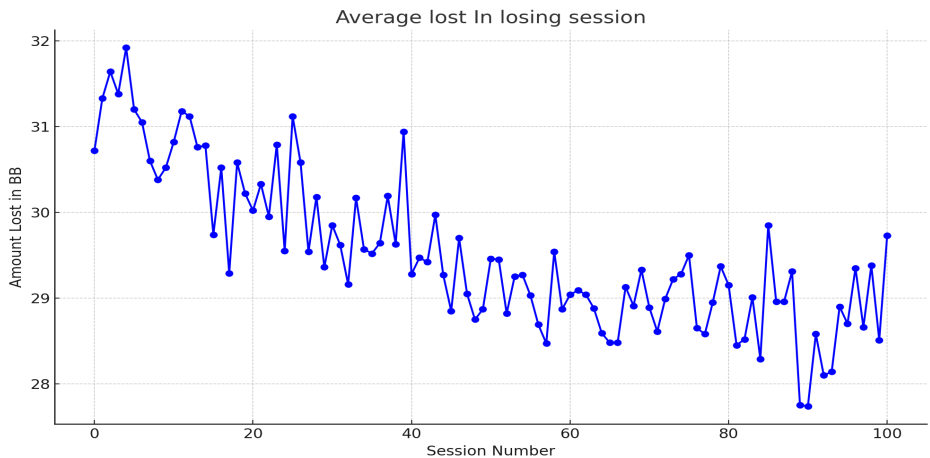

**Figure 3:** Average big blind amount lost by the user vs number of games played by user in 6P Scenario

To get valuable insights into the playing style of experienced users, showcasing their preference for a tighter gameplay approach, characterized by selective hand choices, we plotted tightness versus the number of hands played and the results are given in Figure 4. The X-axis represents the experience of the user, indicated as a percentile based on the number of hands played. The Y-axis represents the tightness of play, reflecting the user's tendency to be more cautious and disciplined in their hand selection. It is important to note that only users who have played at least 100 hands on the platform were considered for this analysis.

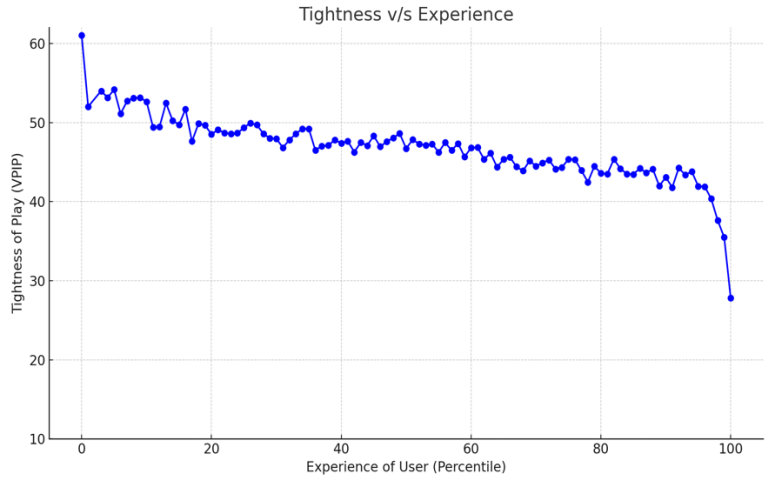

**Figure 4:** Tightness versus number of hands played by user in 6P Scenario

The percentage of winners as the players are playing more and more games is given in Figure 5 in the 6P scenario. It is evident from the plot that the percentage of wins' increases with the number of games played by the user

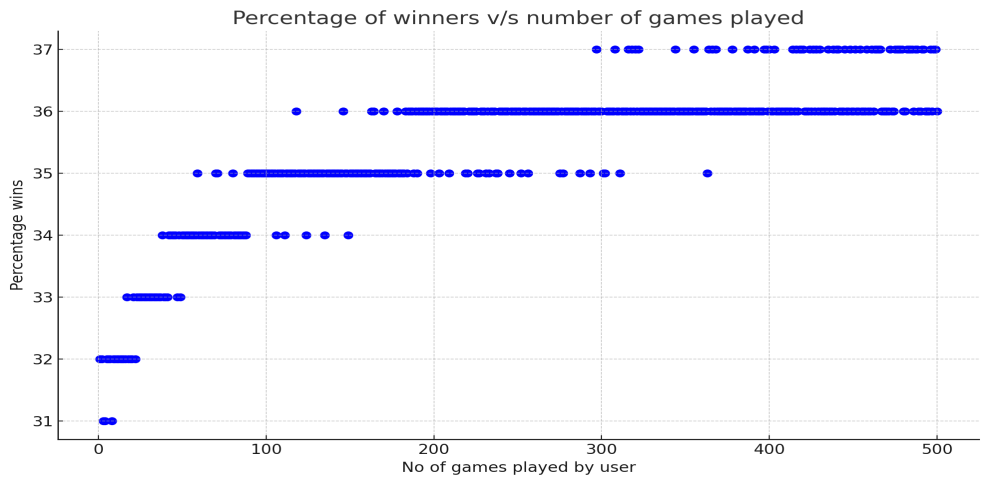

**Figure 5:** Percentage of winner's vs number of games played by the users in 6P Scenario

For an example player who has played over 1000 games, we tried to analyze the winning probability and the results are given in Figure 6. From the diagrammatic illustration it is seen that the winning probability fluctuates between 0.3 to 0.4 with increase in number of games

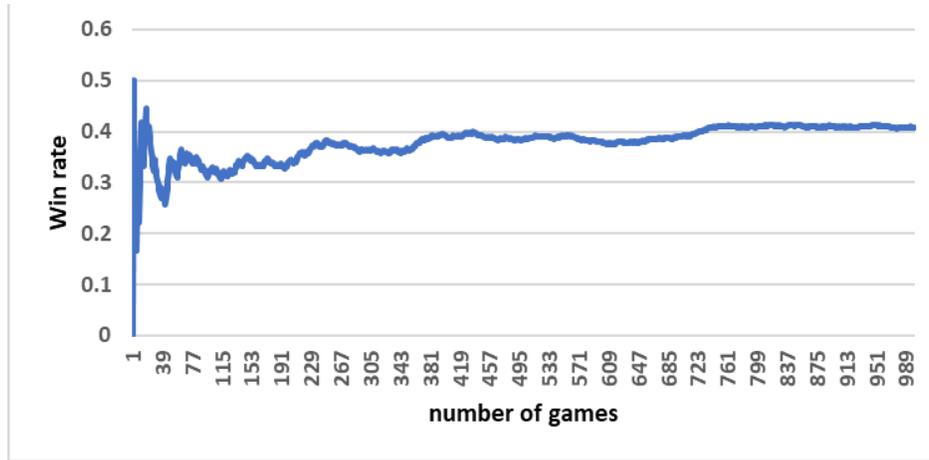

**Figure 6:** Variation in the winning probability for 6-Player scenario in the game of poker

For checking the normality of the distribution of the win percentage, we have used the QQ plots to validate the analysis in each quintile class. Figure 7 shows the Q-Q plot comparing the distribution of the winning percentages over the entire quintile against the normal distribution. The Q-Q plot is a graph between theoretical quintile and the observed value. The deviations from the normal indicated in the plots shows a non-linear pattern suggesting that the mean of the data is not 0 indicating a skill factor involved.

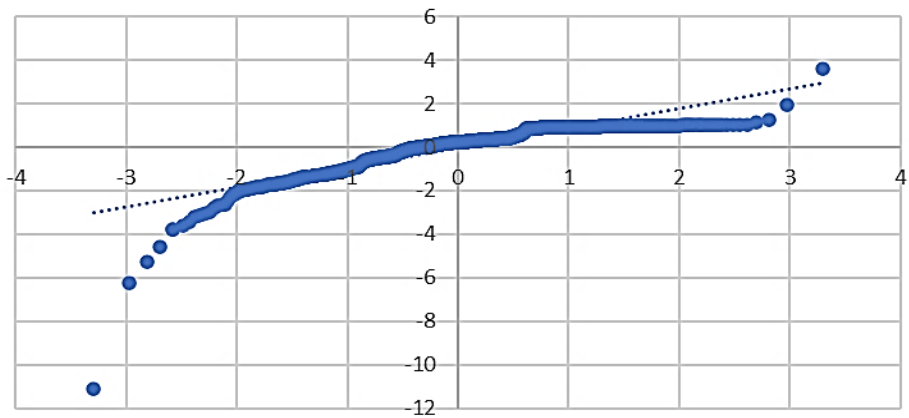

**Figure 7:** Q-Q Plot for a skilled player corresponding to winning percentage in a 6P scenario

4.2 Rummy

In the game of rummy, we computed the correlation coefficient for different skill variables for 10,000 users who played at least 30/100 games over a span of one month and the results are given in Table II. The findings indicate that the Average point lost in losing hands came out to have the highest correlation across periods. Additionally, overall correlation coefficient seems to show much better results when a player has played at least 100 games in both periods.

Table II Correlation coefficient for 1000 players who played at least 30/100 games

| Skill Variable | Correlation (Min 30 Games) | Correlation (Min 100 Games) |
|---|---|---|
| Avg. Points Lost by opponent | 0.32 | 0.53 |
| Win Rate | 0.14 | 0.23 |
| Avg. Points lost in losing hands | 0.64 | 0.77 |

Also, in the game of rummy as the players play more and more games, they become skilled at picking and discarding cards. They also start reading into opponents' strategy which further helps them in making decisions. All this leads to an increase in their win rate as shown in Figure 8.

As the player matures, he becomes more adroit at drop and pick which ensures that even if he loses his loss is minimized. Average points lost in losing hands comes down for the players over a period and the same is being schematically shown in Figure 9.

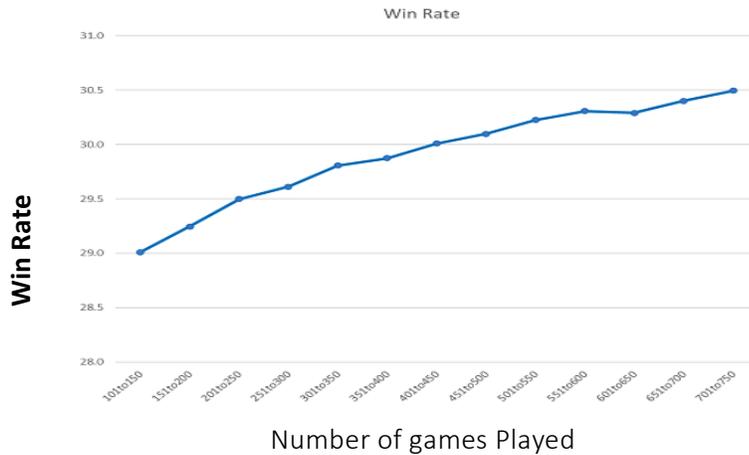

**Figure 8:** Win Rate Versus the number of games played by the user in 6P scenario

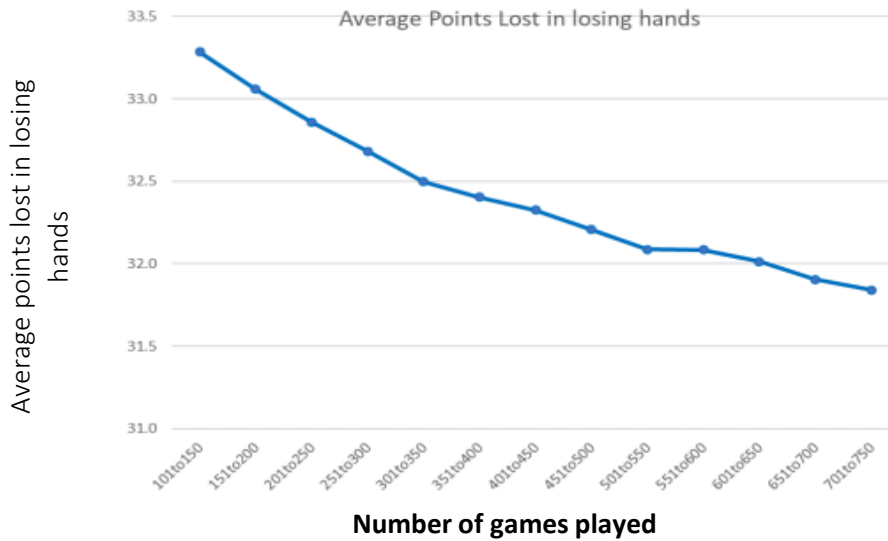

**Figure 9:** Average Points lost in the losing hands Versus the number of games played by the user in 6P scenario

We also investigated the performance of the individual players in terms of the probability of winning as they played more and more games. We analyzed all the three different scenarios, i.e., 2 Player, 3 Player, and 6 player cases and the results are shown in Figure 10. In a 2-player scenario, the win rate first decreases for the first few games, then increases significantly and then shows consistency. The performance of the player becomes constant for the rest (and major part) of the time period considered. A similar trend was also observed in a 3-player scenario. For a 6-player scenario, the player exhibited a fluctuating win rate for the first few games, and then reached almost a constant value.

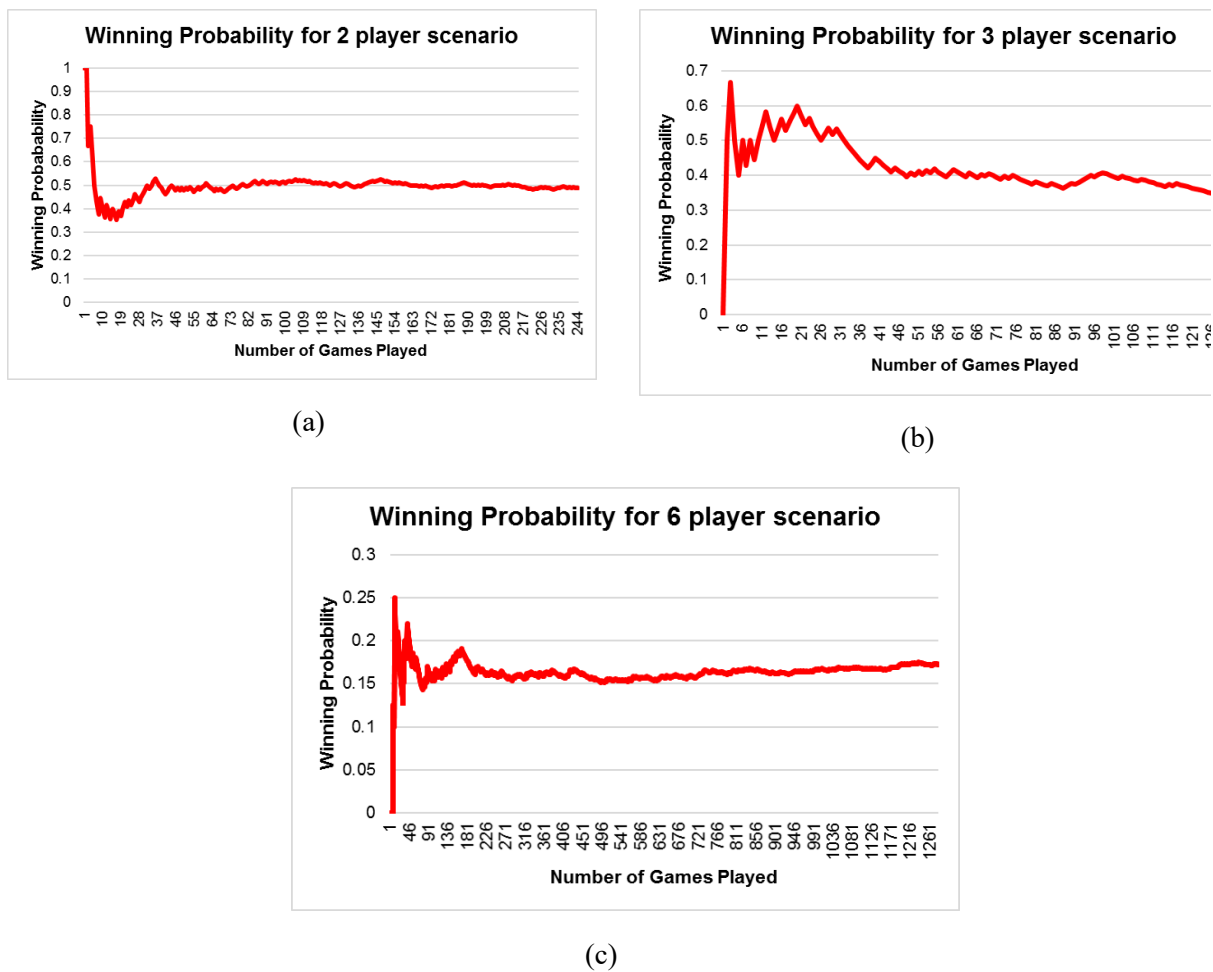

**Figure 10:** Variation in the winning probability for (a) 2-Player scenario, (b) 3-Player scenario, and 6-Player scenario

For checking the normality of the distribution of the win percentage, we have used the QQ plots to validate the analysis in the game of rummy as well. Figure 11 shows the Q-Q plot comparing the distribution of the winning percentages against the normal distribution. Clearly, the deviances from normal indicates that the win rate in long terms is not a random chance but attributed to skill factor

The deviations from the normal indicated in the plots show a non-linear pattern suggesting that the mean of the data is not 0. This also implies that the long-term success in the games played is attributed to skill and not randomness.

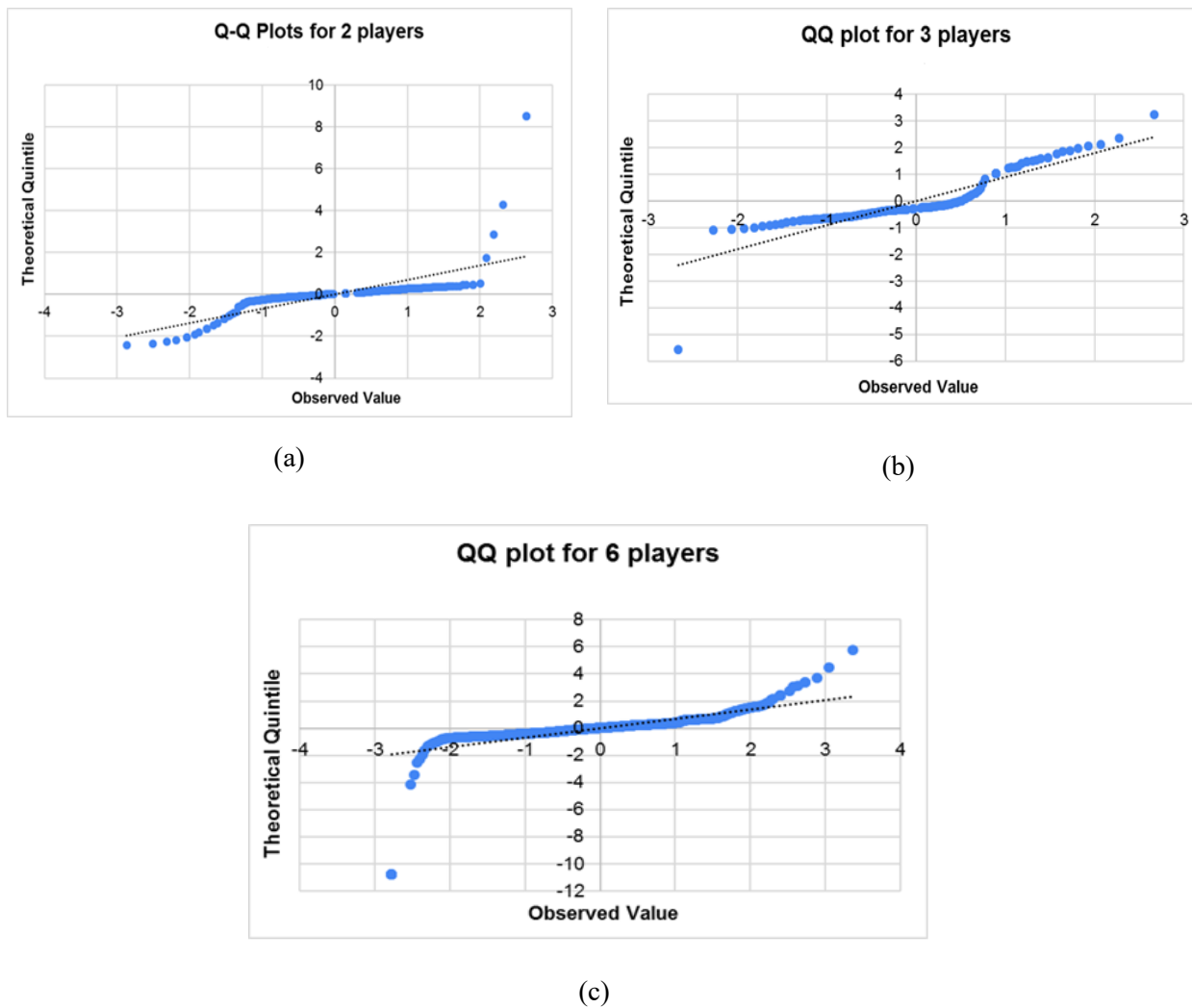

**Figure 11:** Q-Q plots corresponding to the winning percentage in (a) 2-Player scenario, (b) 3-Player scenario, and 6-Player scenario

## 5. Discussion

We also contrasted the mean and standard deviation of the different percentiles by dividing the data into ten/four quartiles in case of poker and rummy. We observe that the mean win rate of the quartiles shows a similar trend as of the overall win rates (indicative in Table III-VI). The tabular findings also illustrate that the standard deviations of the quartiles keep decreasing over the number of quartiles, indicating a decrease in the variability of the win rates. This further strengthens our assumption that the game of poker and rummy are not purely based on chance but has a skill factor involved.

From a neuroscientific perspective, the variation of winning rate of players as they play more games can be studied under the framework of the power law of practice [26]. As per the law, the performance of a player in a new task would show larger improvements in the earlier stages of gameplay and slowly saturate to an asymptotic value. This type of trajectory is linked to the acquisition of skill in the particular task. This trend was also observed in the data from online poker and rummy, suggesting a possible involvement of skill in the players' long-term performance. Moreover, it is interesting to observe that despite the lack of social information cues from the opponents, it is possible to improve one's performance in card games, possible by improvement in visual working memory capacity (WMC). A study by Meinz *et al.* showed that WMC was an important predictor of performance in poker irrespective of domain knowledge in the game [22].

Table III Summary of the aggregated data in 6 Player –scenario for game of poker

| Percentile | Number of players | Mean of win rate | Standard Deviation |
|---|---|---|---|
| 1 | 100 | 0.3273 | 0.0569 |
| 2 | 200 | 0.332 | 0.0102 |
| 3 | 300 | 0.3662 | 0.0114 |
| 4 | 400 | 0.3721 | 0.0128 |
| 5 | 500 | 0.3893 | 0.0044 |
| 6 | 600 | 0.3871 | 0.0037 |
| 7 | 700 | 0.382 | 0.0043 |
| 8 | 800 | 0.4047 | 0.0071 |
| 9 | 900 | 0.4106 | 0.0013 |
| 10 | 1000 | 0.409 | 0.0016 |

Table IV Summary of the aggregated data in 2 Player –scenario for game of rummy

| Quartile | Number of players | Mean | Standard deviation |
|---|---|---|---|
| 1 | 62 | 0.4842 | 0.1166 |
| 2 | 123 | 0.5048 | 0.0132 |
| 3 | 183 | 0.5033 | 0.0089 |
| 4 | 244 | 0.4953 | 0.0072 |

Table V Summary of the aggregated data in 3 Player –scenario for game of rummy

| Quartile | Number of players | Mean | Standard Deviation |
|---|---|---|---|
| 1 | 33 | 0.5096 | 0.10394 |
| 2 | 65 | 0.4174 | 0.01961 |
| 3 | 97 | 0.3863 | 0.01801 |
| 4 | 130 | 0.37395 | 0.01209 |

Table VI Summary of the aggregated data in 6 Player –scenario game of rummy

| Quartile | Number of players | Mean | Standard deviation |
|---|---|---|---|
| 1 | 321 | 0.1633 | 0.1166 |
| 2 | 641 | 0.1581 | 0.0132 |
| 3 | 961 | 0.1624 | 0.0089 |
| 4 | 1281 | 0.1687 | 0.0072 |

## 6. Conclusion

The present work provides a quantitative analysis via simple mathematical tools to test whether the long-term success in online games of poker and rummy could be attributed to chance or skill. Our data for analysis was a set of users who have played a minimum of 30 games and a maximum of 100 games in 2 players, 3 players, and 6 player table. We found that both in the game of poker and rummy, the skill variables improve as the users play more and more games. Upon checking the normality of the winning percentage, we could see that there is a deviation from the normal distribution signifying a non-linear trend between the theoretical and the observed quintiles. This is a precursor that the mean of the winning rate as the user play's more and more games is not zero, which implies that the long-term success cannot be considered a random phenomenon. This observation not only underscores the significance of the deviation but also lends weight to the argument that the outcomes aren't purely products of random chance, but rather are influenced by the players' skills and expertise. The standard deviation of the winning percentage also high lights the performance of the player as they play a greater number of games becomes consistent and the variability gets reduced. From the studies undertaken, we conclude that 1) there is no difference in online and offline versions of rummy and poker from the perspective of requirement of skills and 2) In online versions of rummy and poker, there is preponderance of skills over chance to succeed.

## Acknowledgement

We would like to thank the All India Gaming Federation (AIGF) for sharing the anonymous data of Rummy and Poker for the above analysis with due approval (Ref: Letter dated 23rd Aug, 2023).

# References

[1] M. D. Griffiths, M. N. O. Davies, and D. Chappell, "Demographic factors and playing variables in online computer gaming," *CyberPsychology Behav.*, vol. 7, no. 4, pp. 479–487, 2004.

[2] D. Columb, M. D. Griffiths, and C. O'Gara, "Online gaming and gaming disorder: More than just a trivial pursuit," *Ir. J. Psychol. Med.*, vol. 39, no. 1, pp. 1–7, 2022.

[3] M. D. Griffiths, M. N. O. Davies, and D. Chappell, "Online computer gaming: a comparison of adolescent and adult gamers," *J. Adolesc.*, vol. 27, no. 1, pp. 87–96, 2004.

[4] R. Kandel Eric, H. Schwartz James, and M. Jessell Thomas, "Principles of neural science. Vol. 4." New York: McGraw-hill, 2000.

[5] K. VanLehn, "Cognitive skill acquisition," *Annu. Rev. Psychol.*, vol. 47, no. 1, pp. 513–539, 1996.

[6] D. J. McFarland, "How neuroscience can inform the study of individual differences in cognitive abilities," *Rev. Neurosci.*, vol. 28, no. 4, pp. 343–362, May 2017, doi: 10.1515/revneuro-2016-0073.

[7] E. E. Matthews, T. Brown, and K. Stagnitti, "Relationship Between Sensory Processing and Perceptions of and Participation in Play and Leisure Activities Among Typically Developing Children: An Exploratory Study," *Ann. Int. Occup. Ther.*, vol. 4, no. 2, pp. 85–92, 2021.

[8] M. Ashford, A. Abraham, and J. Poolton, "Understanding a player's decision-making process in team sports: A systematic review of empirical evidence," *Sports*, vol. 9, no. 5, p. 65, 2021.

[9] S. Kairouz, C. Paradis, and L. Nadeau, "Are online gamblers more at risk than offline gamblers?," *Cyberpsychology, Behav. Soc. Netw.*, vol. 15, no. 3, pp. 175–180, 2012.

[10] T. R. K. Motwani, "Skill versus Chance-The Gambling Debate," *Supreme Court Cases J.*, vol. 7, p. 26, 2014.

[11] R. J. D. Potter van Loon, M. J. van den Assem, and D. van Dolder, "Beyond Chance? The Persistence of Performance in Online Poker," *PLoS One*, vol. 10, no. 3, p. e0115479, Mar. 2015, doi: 10.1371/journal.pone.0115479.

[12] I. C. Fiedler and J.-P. Rock, "Quantifying Skill in Games—Theory and Empirical Evidence for Poker," *Gaming Law Rev. Econ.*, vol. 13, no. 1, pp. 50–57, Feb. 2009, doi: 10.1089/glre.2008.13106.

[13] S. D. Levitt, T. J. Miles, and A. M. Rosenfield, "Is Texas Hold'Em a Game of Chance-A Legal and Economic Analysis," *Geo. LJ*, vol. 101, p. 581, 2012.

[14] N. Alon, "Poker, chance and skill," *Unpubl. Manuscr.*, pp. 1–17, 2007.

[15] G. Meyer, M. von Meduna, T. Brosowski, and T. Hayer, "Is Poker a Game of Skill or Chance? A Quasi-Experimental Study," *J. Gambl. Stud.*, vol. 29, no. 3, pp. 535–550, Sep. 2013, doi: 10.1007/s10899-012-9327-8.

[16] N. E. Turner, "Poker is an acquired skill," *Gaming L. Rev. Econ.*, vol. 12, p. 229, 2008.

[17] H. Johansen-Berg and V. Walsh, "Cognitive neuroscience: who to play at poker," *Curr. Biol.*, vol. 11, no. 7, pp. R261--R263, 2001.

[18] M.-A. Vanderhasselt, S. Kühn, and R. De Raedt, "'Put on your poker face': neural systems supporting the anticipation for expressive suppression and cognitive reappraisal," *Soc. Cogn. Affect. Neurosci.*, vol. 8, no. 8, pp. 903–910, Dec. 2013, doi: 10.1093/scan/nss090.

[19] E. Hurel *et al.*, "Spatial attention to social information in poker: A neuropsychological study using the Posner cueing paradigm," *J. Behav. Addict.*, vol. 12, no. 1, pp. 219–229, Mar. 2023, doi: 10.1556/2006.2022.00082.

[20] M. W. Eysenck and M. T. Keane, *Cognitive psychology: A student's handbook*. Taylor & Francis, 2005.

[21] M. Schiavella, M. Pelagatti, J. Westin, G. Lepore, and P. Cherubini, "Profiling Online Poker Players: Are Executive Functions Correlated with Poker Ability and Problem Gambling?," *J. Gambl. Stud.*, vol. 34, no. 3, pp. 823–851, Sep. 2018, doi: 10.1007/s10899-017-9741-z.

[22] E. J. Meinz, D. Z. Hambrick, C. B. Hawkins, A. K. Gillings, B. E. Meyer, and J. L. Schneider, "Roles of domain knowledge and working memory capacity in components of skill in Texas Hold'Em poker.," *J. Appl. Res. Mem. Cogn.*, vol. 1, no. 1, pp. 34–40, Mar. 2012, doi:

[23] K. Gielis *et al.*, "Dissecting Digital Card Games to Yield Digital Biomarkers for the Assessment of Mild Cognitive Impairment: Methodological Approach and Exploratory Study," *JMIR Serious Games*, vol. 9, no. 4, p. e18359, Nov. 2021, doi: 10.2196/18359.
[24] C.-Y. Kuo, Y.-M. Huang, and Y.-Y. Yeh, "Let's Play Cards: Multi-Component Cognitive Training With Social Engagement Enhances Executive Control in Older Adults," *Front. Psychol.*, vol. 9, Dec. 2018, doi: 10.3389/fpsyg.2018.02482.
[25] T. Stafford, S. Devlin, R. Sifa, and A. Drachen, "Exploration and skill acquisition in a major online game," 2017.
[26] G. S. Snoddy, "Learning and stability: a psychophysiological analysis of a case of motor learning with clinical applications.," *J. Appl. Psychol.*, vol. 10, no. 1, pp. 1–36, Mar. 1926, doi: 10.1037/h0075814.
[27] A. Heathcote, S. Brown, and D. J. K. Mewhort, "The power law repealed: The case for an exponential law of practice," *Psychon. Bull. Rev.*, vol. 7, no. 2, pp. 185–207, Jun. 2000, doi: 10.3758/BF03212979.
[28] M. Ministry of Electronics and Information Technology, "Gazette Notification for IT Amendment Rules, 2023," *The Gazette of India*, 2023.
[29] A. Moreau, H. Chabrol, and E. Chauchard, "Psychopathology of online poker players: Review of literature," *J. Behav. Addict.*, vol. 5, no. 2, pp. 155–168, 2016, doi: 10.1556/2006.5.2016.035.
[30] T. L. MacKay, N. Bard, M. Bowling, and D. C. Hodgins, "Do pokers players know how good they are? Accuracy of poker skill estimation in online and offline players," *Comput. Human Behav.*, vol. 31, no. 1, pp. 419–424, 2014, doi: 10.1016/J.CHB.2013.11.006.
[31] J. Hergueux and G. Smagghue, "The dominance of skill in online poker," *Int. Rev. Law Econ.*, vol. 74, p. 106119, 2023, doi: 10.1016/J.IRLE.2022.106119.
[32] T. Mukherjee and S. Eswaran, "Towards mining of player intent for targeted gaming services," *Proc. - 2018 IEEE World Congr. Serv. Serv. 2018*, pp. 45–46, 2018, doi: 10.1109/SERVICES.2018.00041.
[33] A. K. SINGH, "Laws on Online Gaming and Online Gambling in India: Future Market of India," *SSRN Electron. Journa*, 2023, doi: https://dx.doi.org/10.2139/ssrn.4480470.
[34] https://www.gamblingcommission.gov.uk/print/remote-gambling-and-software-technical-standards